\documentclass[11pt]{article}
\usepackage[dvips]{color}
\usepackage{epsfig}
\usepackage{amsmath}
\usepackage{graphicx}
\date{}
\begin{document}
\title{{\bf A late time accelerated FRW model with scalar and vector fields via Noether symmetry}}
\author{Babak Vakili\thanks{b-vakili@iauc.ac.ir}\\\\{\small {\it Department of Physics, Central Tehran Branch, Islamic Azad University, Tehran,
Iran}}} \maketitle

\begin{abstract}
We study the evolution of a three-dimensional minisuperspace
cosmological model by the Noether symmetry approach. The phase space
variables turn out to correspond to the scale factor of a flat
Friedmann-Robertson-Walker (FRW) model, a scalar field with
potential function $V(\phi)$ with which the gravity part of the
action is minimally coupled and a vector field of its kinetic energy
is coupled with the scalar field by a coupling function $f(\phi)$.
Then, the Noether symmetry of such a cosmological model is
investigated by utilizing the behavior of the corresponding
Lagrangian under the infinitesimal generator of the desired
symmetry. We explicitly calculate the form of the coupling function
between the scalar and the vector fields and also the scalar field
potential function for which such symmetry exists. Finally, by means
of the corresponding Noether current, we integrate the equations of
motion and obtain exact solutions for the scale factor, scalar and
vector fields. It is shown that the resulting cosmology is an
accelerated expansion universe for which its expansion is due to the
presence of the vector field in the early times, while the scalar
field is responsible of its late time expansion.
\vspace{5mm}\noindent\vspace{0.8mm}\newline Keywords:
Noether symmetry, Scalar field cosmology, Vector field cosmology
\end{abstract}

\section{Introduction}
Symmetries have always played a central role in conceptual
discussion of the classical and quantum physics. The main reason may
be that various laws of conservation, such as energy, momentum,
angular momentum, etc., that provide the integrals of motion for a
given dynamical system, are indeed the result of existence of some
kinds of symmetry in that system. From a more general point of view,
it can be shown that all such conservation laws are particular cases
of the so-called Noether theorem, according to which for every
one-parameter group of coordinate transformation on the
configuration space of a system, which preserves the Lagrangian
function, there exists a first integral of motion \cite{Arnold}. In
mathematical language this means that if the vector field $X$ is the
generator of the above diffeomorphism, the Lie derivative of the
Lagrangian function along it should vanish: $L_X{\cal L}=0$
\cite{Capp1}. Numerous applications of this theorem in general
relativity and cosmology are those concerned with the following form
of action (see for instance \cite{app} and the references therein)
\begin{equation}\label{Int1}
{\cal S}=\int_{{\cal M}} d\tau \left[\frac{1}{2}{\cal
G}_{AB}\frac{dq^A}{d\tau}\frac{dq^B}{d\tau}-{\cal U}(\bf
{q})\right],\end{equation}where $q^A$ are the coordinates of the
configuration space with metric ${\cal G}_{AB}$ (the indices $A$,
$B$, ... run over the dimension of this space), ${\cal U}(\bf{q})$
is the potential function and $\tau$ is an affine parameter along
the evolution path of the system. In time-parameterized theories
such as general relativity, the action retains its form under time
reparameterization. Therefore, one may relate the affine parameter
$\tau$ to a time parameter $t$ by a lapse function $N(t)$ through
$Ndt=d\tau$. In these cases the action (\ref{Int1}) can be written
as
\begin{equation}\label{Int2}
{\cal S}=\int_{{\cal M}}dt{\cal L}(q^A,\dot{q}^A)=\int_{{\cal M}}dt
N\left[\frac{1}{2N^2}{\cal G}_{AB}\dot{q}^A\dot{q}^B-{\cal
U}(\bf{q})\right],\end{equation}where an over-dot indicates
derivation with respect to the time parameter $t$ and ${\cal
L}(\bf{q},\dot{\bf{q}})$ is the Lagrangian function of the system. A
straightforward calculation based on the Hamiltonian formalism leads
us to the Hamiltonian constraint
\begin{equation}\label{Int3}
H=N\left[\frac{1}{2}{\cal G}^{AB}P_AP_B+{\cal
U}(\bf{q})\right]=N{\cal H}\equiv 0,\end{equation}where $P_A$ is the
momentum conjugate to $q^A$.

For the systems whose dynamics can be described by the above
explanation, we may define a vector field $X$ on the tangent space
$TQ=({\bf q}, {\bf \dot{q}})$ by
\begin{equation}\label{Int4}
X=\alpha^A({\bf q})\frac{\partial}{\partial
q^A}+\frac{d\alpha^A({\bf q})}{dt}\frac{\partial}{\partial
\dot{q}^A},\end{equation}where $\alpha^A({\bf q})$ are unknown
functions on configuration space. According to what we have
mentioned above, this vector field will generate a Noether symmetry
if
\begin{equation}\label{Int5}
L_X{\cal L}=\alpha^A({\bf q})\frac{\partial {\cal L}}{\partial
q^A}+\frac{d\alpha^A({\bf q})}{dt}\frac{\partial {\cal L}}{\partial
\dot{q}^A}=0.\end{equation}As the other kinds of symmetries, the
Noether symmetry is also a powerful tool in finding the solutions of
equations of motion in a dynamical system. Indeed, as we will see in
detail in the following sections, noting that $P_A=\frac{\partial
{\cal L}}{\partial \dot{q}^A}$ and taking into account the
Euler-Lagrange equations $\frac{dP_A}{dt}=\frac{\partial {\cal
L}}{\partial q^A}$, from (\ref{Int5}) we get the first integrals of
motion as
\begin{equation}\label{Int6}
Q=\alpha^A({\bf q})P_A.\end{equation}

In cosmology, when the models are expressed in terms of the
minisuperspace variables usually the scale factors, matter fields
and their conjugate momenta play the role of dynamical variables. In
these models it can be shown that the evolution of the system can be
obtained from an action of the form (\ref{Int2}) \cite{Hall}.
Therefore, introduction of Noether symmetry by adopting the approach
discussed above is particularly relevant. This method usually works
in such a way that first one sets up an effective Lagrangian in
terms of its configuration variables and their velocities so that
its variation yields the appropriate equations of motion. However,
in many of the extended theories of gravity the Lagrangian involves
potential and coupling functions that are not clearly defined. The
functional forms of these functions may have their roots in the
other physical theories such as particle physics and quantum field
theory. In the Noether symmetry approach, the form of the unknown
functions in the Lagrangian may be found by demanding that the
Lagrangian admits the desired Noether symmetry. In this regard, the
condition (\ref{Int5}) gives a system of partial differential
equations from its solutions of the unknown functions $\alpha^A({\bf
q})$ as well as the potential and other coupling functions in the
Lagrangian are extracted.

Our goal in this paper is to explore the Noether symmetry in a
cosmological model for which, in addition to a scalar field, a
vector field is also present in its action. Although scalar fields
have played an important role in the development of modern
cosmological theories \cite{Scalar}, the vector fields to introduce
the various cosmological aspects have seldom been studied in the
literature \cite{Vector}. Our study is based on an action introduced
in \cite{Sheikh} to investigate the anisotropic inflation with gauge
fields in which a scalar field has either a minimally coupling with
gravity or a non-minimally coupling with a vector field, see the
action (\ref{B}) below. However, they have shown that for special
exponential forms (fixed by hand) for the scalar field potential and
the coupling function between the scalar and vector fields, the
model has also isotropic power-law inflationary solutions. This was
a motivation for us to consider the existence of Noether symmetry in
such models with unknown potential and coupling functions. So, we
will consider a flat FRW cosmology with scale factor $a$, a scalar
field $\phi$ with potential $V(\phi)$ minimally coupled to it and a
vector field $A_{\mu}$ non-minimally coupled to the scalar field by
a coupling function $f(\phi)$. Therefore, the corresponding
minisuperspace of our model is a three-dimensional Riemannian
manifold with coordinates $(a,\phi,A)$ in which we construct a
point-like Lagrangian to produce the dynamics of the model. We then
impose the Noether symmetry condition on this Lagrangian and see how
one can obtain the explicit form of the potential and the coupling
functions. Since the existence of a symmetry results in a constant
of motion, we can integrate the field equations which would then
lead to the expansion law of the universe.
\section{The model}
In this section we consider a homogeneous and isotropic cosmological
model in which the space-time is assumed to be of flat FRW whose
line element can be written

\begin{equation}\label{A}
ds^2=-N^2(t)dt^2+a^2(t)\delta_{ij}dx^idx^j,
\end{equation}where $N(t)$ and $a(t)$ are the lapse function and the
scale factor, respectively. In such a background geometry, we
consider a gravity model whose dynamics is given by the action
\cite{Sheikh}

\begin{equation}\label{B}
{\cal S}=\int d^4x
\sqrt{-g}\left[\frac{1}{2}R-\frac{1}{2}g^{\mu\nu}\partial_{\mu}\phi
\partial_{\nu}\phi-V(\phi)-\frac{1}{4}f(\phi)^2F_{\mu\nu}F^{\mu\nu}\right],
\end{equation}where $\phi(t)$ is a scalar field minimally coupled to
gravity, $V(\phi)$ is its potential and $F_{\mu\nu}$ is the strength
tensor of the vector field $A_{\mu}$ with standard definition
$F_{\mu\nu}=\partial_{\mu}A_{\nu}-\partial_{\nu}A_{\mu}$. As the
action shows, the vector field is coupled to the scalar field
through the coupling function $f(\phi)$. If we introduce a vector
field

\begin{equation}\label{C}
A_{\mu}=\left(0;0,0,A(t)\right),
\end{equation} the non-vanishing components of $F_{\mu\nu}$ are

\begin{equation}\label{D}
F_{03}=-F_{30}=\dot{A}\Rightarrow
F^{03}=-F^{30}=-\frac{1}{N^2}a^{-2}\dot{A},
\end{equation}from which we get
\begin{equation}\label{E}
F_{\mu\nu}F^{\mu\nu}=-\frac{2}{N^2}a^{-2}\dot{A}^2.
\end{equation} With the above results at hand, the action (\ref{B})
can be written in the form ${\cal S}=\int dt {\cal L}({\bf
q},\dot{{\bf q}})$, where ${\bf q}=(a,\phi,A)$ and

\begin{equation}\label{F}
{\cal L}({\bf q},\dot{{\bf
q}})=\frac{1}{N}\left[-3a\dot{a}^2+\frac{1}{2}a^3\dot{\phi}^2+\frac{1}{2}af(\phi)^2
\dot{A}^2-N^2a^3V(\phi)\right],
\end{equation}is a point-like Lagrangian from which the dynamics of
the model can be obtained. It is clear that this Lagrangian has the
form of (\ref{Int2}) with
\begin{equation}\label{F1}
{\cal G}_{\mu\nu}=\mbox{diag}(-6a,a^3,af(\phi)^2).
\end{equation}
To write the corresponding Hamiltonian, we notice that the momenta
conjugate to the dynamical variables may be obtained from the
definition $P_q=\frac{\partial {\cal L}}{\partial \dot{q}}$ with
result

\begin{equation}\label{G}
P_a=-\frac{6a\dot{a}}{N},\hspace{5mm}
P_{\phi}=\frac{a^3\dot{\phi}}{N},\hspace{5mm}P_A=\frac{af(\phi)^2\dot{A}}{N},
\end{equation}leading to the following Hamiltonian
\begin{equation}\label{H}
H=N{\cal
H}=N\left[-\frac{P_a^2}{12a}+\frac{P_{\phi}^2}{2a^3}+\frac{P_A^2}{2af(\phi)^2}+a^3V(\phi)\right].
\end{equation}Now, the dynamical equations (in the cosmic time gauge $N=1$) of the system can be
written by using of the Hamiltonian equations that are
\begin{eqnarray}\label{I}
\left\{
\begin{array}{ll}
\dot{a}=\{a, H\}=-\frac{P_a}{6a},\\\\
\dot{P_a}=\{P_a,
H\}=-\frac{P_a^2}{12a^2}+\frac{3P_{\phi}^2}{2a^4}+\frac{P_A^2}{2a^2f(\phi)^2}
-3a^2V(\phi),\\\\
\dot{\phi}=\{\phi,H\}=\frac{P_{\phi}}{a^3},\\\\
\dot{P_{\phi}}=\{P_{\phi},H\}=\frac{P_A^2f'(\phi)}{a f(\phi)^{3}}-a^3V'(\phi),\\\\
\dot{A}=\{A,H\}=\frac{P_A}{a f(\phi)^{2}},\\\\
\dot{P_A}=\{P_A,H\}=0.
\end{array}
\right.
\end{eqnarray}Also, it is well known that the Hamiltonian of a
gravitational system is constrained to vanish due to the invariant
property of the action under time reparameterization. By using of
the relations (\ref{G}) and (\ref{H}), the Hamiltonian constraint
$H=0$ reads

\begin{equation}\label{J}
-3\dot{a}^2+\frac{1}{2}a^2\dot{\phi}^2+\frac{1}{2}f(\phi)^2\dot{A}^2+a^2V(\phi)=0.
\end{equation}It is clear that to solve the above equations first
of all one should decide for the form of the potential function
$V(\phi)$ and the coupling function $f(\phi)$. In the next section
we will fix this issue by demanding that the Lagrangian (\ref{F})
satisfies a Noether symmetry condition.

\section{Noether symmetry}
In this section we assume that the Lie derivative of the Lagrangian
(\ref{F}) along a vector field $X$ vanishes which means that the
model has the so-called Noether symmetry. Under this condition we
have
\begin{equation}\label{K}
L_X {\cal L}=0,\end{equation} where $X$ has the form of equation
(\ref{Int4}), that is

\begin{equation}\label{L}
X=\alpha\frac{\partial}{\partial a}+\beta\frac{\partial}{\partial
\phi}+\gamma\frac{\partial}{\partial
A}+\dot{\alpha}\frac{\partial}{\partial
\dot{a}}+\dot{\beta}\frac{\partial}{\partial
\dot{\phi}}+\dot{\gamma}\frac{\partial}{\partial \dot{A}},
\end{equation}in which $\alpha (a,\phi,A)$, $\beta (a,\phi,A)$ and $\gamma (a,\phi,A)$ are some unknown functions
of the configuration space variables $(a,\phi,A)$. Now, by imposing
the condition (\ref{K}) we arrive at

\begin{eqnarray}\label{M}
0&=&-3\dot{a}^2\left(\alpha+2a\frac{\partial \alpha}{\partial
a}\right)+a^2\dot{\phi}^2\left(\frac{3}{2}\alpha+a\frac{\partial
\beta}{\partial \phi}\right)+f(\phi)\dot{A}^2\left[\frac{1}{2}\alpha
f(\phi)+a\beta f'(\phi)+a^2f(\phi)\frac{\partial \gamma}{\partial
A}\right]\\\nonumber &+& a\dot{a}\dot{\phi}\left(-6\frac{\partial
\alpha}{\partial \phi}+a^2\frac{\partial \beta}{\partial
a}\right)+a\dot{a}\dot{A}\left(-6\frac{\partial \alpha}{\partial
A}+f(\phi)^2\frac{\partial \gamma}{\partial
a}\right)+a\dot{\phi}\dot{A}\left(a^2\frac{\partial \beta}{\partial
A}+f(\phi)^2\frac{\partial \gamma}{\partial \phi}\right)\\\nonumber
&-& a^2\left[3\alpha V(\phi)+a\beta V'(\phi)\right].
\end{eqnarray}It is seen that the above expression is a quadratic polynomial in
terms of $\dot{a}$, $\dot{\phi}$, $\dot{A}$. Therefore, the
necessary and sufficient condition for this expression to be
identically equal to zero is that all of its coefficients are zero
which leads to a system of partial differential equations for
$\alpha$, $\beta$ and $\gamma$. With this argument we are led to the
following system:

\begin{eqnarray}\label{N}
\left\{
\begin{array}{ll}
\alpha+2a\frac{\partial \alpha}{\partial a}=0,\\\\
3\alpha+2a\frac{\partial \beta}{\partial \phi}=0,\\\\
-6\frac{\partial \alpha}{\partial \phi}+a^2\frac{\partial \beta}{\partial a}=0,\\\\
-6\frac{\partial \alpha}{\partial A}+f(\phi)^2\frac{\partial \gamma}{\partial a}=0,\\\\
a^2\frac{\partial \beta}{\partial A}+f(\phi)^2\frac{\partial \gamma}{\partial \phi}=0,\\\\
\alpha f(\phi)+2a\beta f'(\phi)+2a^2f(\phi)\frac{\partial \gamma}{\partial A}=0,\\\\
3\alpha V(\phi)+a\beta V'(\phi)=0.
\end{array}
\right.
\end{eqnarray}From the first equation of this
system, we can immediately separate the function $\alpha$ as

\begin{equation}\label{O}
\alpha(a,\phi,A)=a^{-1/2}G(\phi)H(A),
\end{equation}where $G(\phi)$
and $H(A)$ are arbitrary functions of $\phi$ and $A$, respectively.
Using this expression in the second equation of the system (\ref{N})
we obtain

\begin{equation}\label{P}
\beta(a,\phi,A)=-\frac{3}{2}a^{-3/2}H(A)\int G(\phi)d\phi.
\end{equation}Upon substitution these results into the third
equation of (\ref{N}) we obtain the following expression for
$G(\phi)$:
\begin{equation}\label{Q}
G(\phi)=c_1e^{\omega \phi}+c_2e^{-\omega \phi},
\end{equation}where $\omega^2=3/8$ and $c_{1,2}$ are integration
constants. Now from the fourth equation of (\ref{N}) we have
$\frac{\partial \gamma}{\partial
a}=6a^{-1/2}f^{-2}(\phi)G(\phi)\frac{dH}{dA}$ which after
integration with respect to $a$ gives

\begin{equation}\label{R}
\gamma(a,\phi,A)=12a^{1/2}f^{-2}(\phi)G(\phi)\frac{dH}{dA}.
\end{equation}Computing $\frac{\partial \gamma}{\partial \phi}$ from
this equation and equating it with $\frac{\partial \gamma}{\partial
\phi}$ from the fifth equation of (\ref{N}) yields

\begin{equation}\label{S}
f(\phi)=G^{1/3}(\phi)=\left(c_1e^{\omega \phi}+c_2e^{-\omega
\phi}\right)^{1/3}.
\end{equation}Using the above relations in the last equation of the
system (\ref{N}) leads us to

\begin{equation}\label{T}
\frac{V'(\phi)}{V(\phi)}=2\omega \frac{c_1e^{\omega
\phi}+c_2e^{-\omega \phi}}{c_1e^{\omega \phi}-c_2e^{-\omega \phi}},
\end{equation}from which we immediately obtain the form of the
potential function as

\begin{equation}\label{U}
V(\phi)=\left(c_1e^{\omega \phi}-c_2e^{-\omega \phi}\right)^2.
\end{equation}What remains is to examine the solutions obtained
until now into the sixth equation of (\ref{N}). If one does so, gets

\begin{equation}\label{V}
\frac{24}{H(A)}\frac{d^2H}{dA^2}=a^{-3}\left(-f(\phi)^2+\frac{3f(\phi)f'(\phi)}{G(\phi)}\int
G(\phi)d\phi \right).
\end{equation}This equation is consistent if its both sides are
equal to zero. Therefore, from $\frac{24}{H(A)}\frac{d^2H}{dA^2}=0$
we obtain
\begin{equation}\label{X}
H(A)=d_1 A+d_2,
\end{equation}
where $d_1$ and $d_2$ are some integration constants. Also, if we
use the relations (\ref{Q}) and (\ref{S}) in the right-hand side of
(\ref{V}) and then put it equal to zero, we arrive at $c_1c_2=0$.
The analysis of the solutions to the system (\ref{N}) is now
complete and two sets of its solutions are achieved as

$\bullet$ I ($c_1=0$)
\begin{eqnarray}\label{Y}
\left\{
\begin{array}{ll}
\alpha(a,\phi,A)=\alpha_0 a^{-1/2}e^{-\omega \phi}(d_1 A+d_2),\\\\
\beta(a,\phi,A)=\beta_0 a^{-3/2}e^{-\omega \phi}(d_1A+d_2),\\\\
\gamma(a,\phi,A)=\gamma_0 a^{1/2}e^{-\omega \phi/3},\\\\
V(\phi)=V_0e^{-2\omega \phi},\\\\
f(\phi)=f_0e^{-\omega \phi/3},
\end{array}
\right.
\end{eqnarray} and

$\bullet$ II ($c_2=0$)
\begin{eqnarray}\label{Z}
\left\{
\begin{array}{ll}
\alpha(a,\phi,A)=\alpha_0 a^{-1/2}e^{\omega \phi}(d_1 A+d_2),\\\\
\beta(a,\phi,A)=-\beta_0 a^{-3/2}e^{\omega \phi}(d_1A+d_2),\\\\
\gamma(a,\phi,A)=\gamma_0 a^{1/2}e^{\omega \phi/3},\\\\
V(\phi)=V_0e^{2\omega \phi},\\\\
f(\phi)=f_0e^{\omega \phi/3},
\end{array}
\right.
\end{eqnarray}where in each case $V_0$ is a positive constant in terms of which
we have $\alpha_0=V_0^{1/2}$, $\beta_0=\frac{3V_0^{1/2}}{2\omega}$,
$\gamma_0=12V_0^{1/6}d_1$ and $f_0=V_0^{1/6}$. It is seen that the
Noether symmetry fixes the potential function for the scalar field
and the coupling function between the scalar and vector fields in
the form of exponential functions. These are exactly the same as are
chosen in \cite{Sheikh} to get exact power-law isotropic inflation
with the help of the above model.

\section{Cosmological dynamics}
With the functions $V(\phi)$ and $f(\phi)$ at hand, we may look for
the solutions to the equations of motion (\ref{I}) and (\ref{J}).
However, before dealing with this issue, we note that the model has
some constants of motion. One of them, $P_{0A}$, comes from the last
equation of (\ref{I}) and indeed is the reflection of the fact that
$A$ is a cyclic variable in the Lagrangian (\ref{F}). Another
constant of motion arises from the existence of the Noether symmetry
in the model under consideration. To see this, we rewrite equation
(\ref{K}) as

\begin{equation}\label{AB}
L_X{\cal L}=\left(\alpha \frac{\partial {\cal L}}{\partial
a}+\frac{d \alpha}{dt}\frac{\partial {\cal L}}{\partial
\dot{a}}\right)+\left(\beta \frac{\partial {\cal L}}{\partial
\phi}+\frac{d \beta}{dt}\frac{\partial {\cal L}}{\partial
\dot{\phi}}\right)+\left(\gamma \frac{\partial {\cal L}}{\partial
A}+\frac{d \gamma}{dt}\frac{\partial {\cal L}}{\partial
\dot{A}}\right)=0.
\end{equation}
Noting from the Euler-Lagrange equation that $\frac{\partial {\cal
L}}{\partial q}=\frac{dP_q}{dt}$, we have
\begin{equation}\label{AC}
\left(\alpha
\frac{dP_a}{dt}+\frac{d\alpha}{dt}P_a\right)+\left(\beta
\frac{dP_{\phi}}{dt}+\frac{d\beta}{dt}P_{\phi}\right)+\left(\gamma
\frac{dP_A}{dt}+\frac{d\gamma}{dt}P_A\right)=0,
\end{equation}
which yields
\begin{equation}\label{AD}
\frac{d}{dt}\left(\alpha P_a+\beta P_{\phi}+\gamma P_A\right)=0.
\end{equation}
Thus the constant of motion is found as
\begin{equation}\label{AE}
Q=\alpha P_a+\beta P_{\phi}+\gamma P_A.
\end{equation}With the help of the relations (\ref{G}), (\ref{Y})
and (\ref{Z}) the above integral of motion can be written as

\begin{equation}\label{AF}
\left(\alpha_0 a^{-1/2}e^{\mp \omega \phi}A\right)(-6a\dot{a})\pm
\left(\beta_0 a^{-3/2}e^{\mp \omega \phi}
A\right)(a^3\dot{\phi})+\left(\gamma_0 a^{1/2}e^{\mp \omega
\phi/3}\right)(a f_0^2 e^{\mp 2\omega \phi/3}\dot{A})=Q,
\end{equation}in which we have set $d_1=1$ and $d_2=0$. Also, the
upper and lower signs correspond to the class I and class II
solutions, respectively. After a little algebra in which the fifth
equation of the system (\ref{I}) is also considered, this expression
results in

\begin{equation}\label{AG}
-\frac{3}{2}\frac{\dot{a}}{a}\pm \omega \dot{\phi} +
3\frac{\dot{A}}{A}=\frac{1}{4}QP_{0A}^{-3/2}\frac{\dot{A}^{3/2}}{A}.
\end{equation}In what follows we shall deal with the solutions of
the above equation only for the simple case $Q=0$ for which
integration of (\ref{AG}) gives

\begin{equation}\label{AH}
\omega \phi=\mp \ln \frac{A^3}{a^{3/2}}.
\end{equation}With the help of these relations and from the last equations of (\ref{Y})
and (\ref{Z}) the expression $f=f_0A a^{-1/2}$ will be obtained in
which the function $f$ is expressed in terms of $A$ and $a$. By
using of this result in the fifth equation of the system (\ref{I})
we obtain

\begin{equation}\label{AI}
A(t)=\left(\omega_0 t\right)^{1/3},
\end{equation}where $\omega_0=3f_0^{-2}P_{0A}$ and we have set
$P_A=P_{0A}=\mbox{cons.}$ from the last equation of (\ref{I}). Now,
we may insert the relations (\ref{AH}), (\ref{AI}) and the
expressions for $V(\phi)$ from (\ref{Y}) and (\ref{Z}) into
(\ref{J}) to get the following equation for the scale factor

\begin{equation}\label{AJ}
-ta^2\dot{a}+\frac{1}{3}a^3+\frac{1}{9}\xi t^{4/3}+\eta t^4=0,
\end{equation}where $\xi=f_0^2\omega_0^{4/3}/8$ and
$\eta=V_0\omega_0^2/4$. This equation can be easily integrated to
yield

\begin{equation}\label{AL}
a(t)=\left(\eta t^4+\xi t^{4/3}+{\cal C}t\right)^{1/3},
\end{equation}in which ${\cal C}$ is an integration constant.
Finally, the time evolution of the scalar field can now be achieved
from (\ref{AH}) as

\begin{equation}\label{AK}
\omega \phi(t)=\mp \ln \frac{\omega_0 t}{\left(\eta t^4+\xi
t^{4/3}+{\cal C}t\right)^{1/2}}.\end{equation}The above expressions
for the corresponding cosmology which describe an isotropic
accelerated expansion universe, are comparable with the relations
(4.39) and (4.41) of \cite{Sheikh}. Figure 1 shows the qualitative
behavior of the scale factor and scalar field for the class I of
solutions. As is clear from the figure, in the early times of cosmic
evolution, the amount of the scalar field (and also its kinetic
energy) decreases while at the same time, according to relations of
(\ref{Y}), the coupling with the vector field (and also the vector
field's kinetic energy) is growing. So during this period, the
coupling between the scalar and the vector fields is responsible for
the expansion of the universe. However, after the scalar field
reaches its minimum value, its incremental behavior begins. A glance
at the relations of (\ref{Y}) shows that in this era the coupling
function rapidly decreases and the vector field loses its energy.
Therefore, the late time acceleration is due to the presence of the
scalar field without a significant roll of the vector field. A
similar discussion can be raised for the class II of solutions.
\begin{figure}
\includegraphics[width=2.5in]{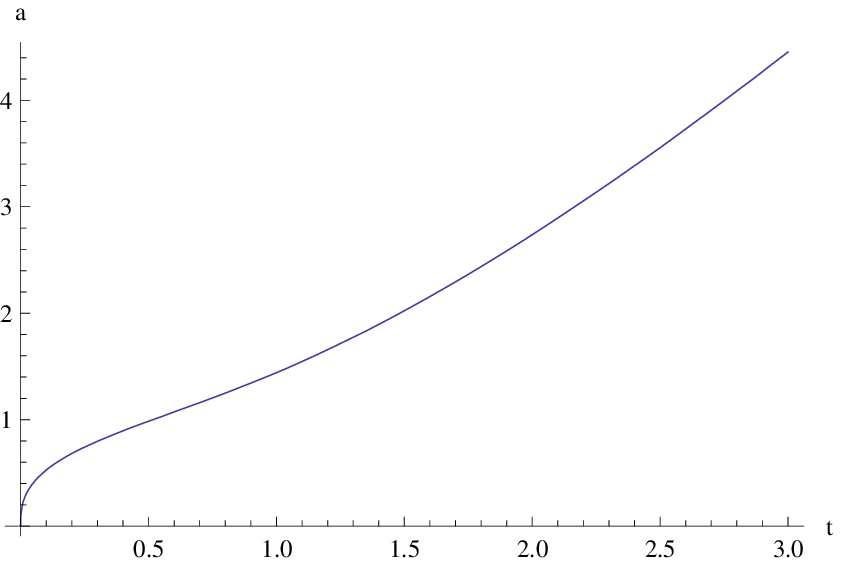}\hspace{3cm}\includegraphics[width=2.5in]{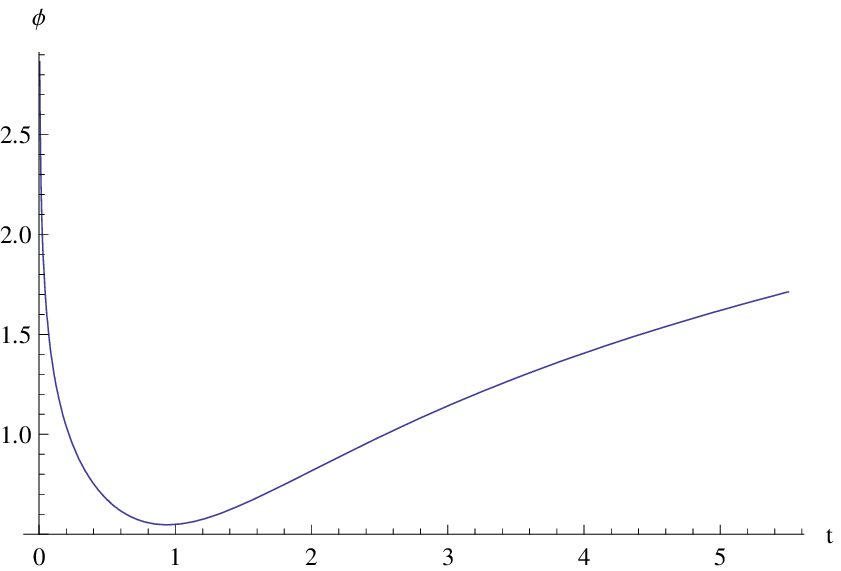}
\caption{Qualitative behavior of the scale factor (left) and scalar
field (right) versus time.}
\end{figure}

\section{Summary}
In this letter, we have studied a scalar-vector field model of
cosmology in a Noether symmetry point of view, in such a way that in
its action, in addition to a minimally coupling between the scalar
field and gravity, there is also a coupling between the scalar and
the kinetic energy of the vector field. For the background geometry,
we have considered a flat FRW metric and then set up the phase space
by taking the scale factor $a$, scalar field $\phi$ and the vector
field $A$ as the independent dynamical variables. The Lagrangian of
the model in the configuration space spanned by $\{a,\phi,A\}$ is so
constructed that its variation with respect to these dynamical
variables yields the Einstein field equations. The existence of
Noether symmetry implies that the Lie derivative of this Lagrangian
with respect to the infinitesimal generator of the desired symmetry
vanishes. By applying this condition to the Lagrangian of the model,
we have obtained the explicit form of the corresponding potential
function of the scalar field and the coupling function between the
scalar and the vector fields. We then obtained the constant of
motion related to the Noether symmetry by means of which we could
integrate the dynamics to yield the exact expressions for the
dynamical variables $a(t)$, $\phi(t)$ and $A(t)$. The evolutionary
behavior of these quantities shows that with a growing vector field
we have an isotropic accelerated expansion universe. Our analysis
showed that in the early times of evolution the amount of the scalar
field decreases continuously to reach a minimum value,
simultaneously the kinetic energy of the vector field and its
coupling to the scalar field increase. On the other hand, after this
period this behavior is reversed, the scalar field begins to
increase while its coupling with vector field as well as the vector
field's kinetic energy rapidly decrease. Therefore, in the late
times, the universe is scalar field dominated and the vector field
plays a subdued role in the expansion in this epoch.

\end{document}